# DOPING EFFECTS IN SINGLE-LAYERED $La_{0.5}Sr_{1.5}MnO_4$ MANGANITES

*Lorenzo Malavasi*[1,*], *Anna Maria Bertasa*[1], *Maria Cristina Mozzati*[2], *Cristina Tealdi*[1], *Carlo Bruno Azzoni*[2], *and Giorgio Flor*[1]

[1]Dipartimento di Chimica Fisica "M. Rolla", INSTM, Università di Pavia, V.le Taramelli 16, I-27100, Pavia, Italy.

[2]CNISM, Unità di Pavia and Dipartimento di Fisica "A. Volta", Università di Pavia, Via Bassi 6, I-27100, Pavia, Italy.

In this paper we report the results of the synthesis and structural, transport and magnetic characterization of pure $La_{0.5}Sr_{1.5}MnO_4$ (LSMO) and 5% doped samples, *i.e.* $La_{0.5}Sr_{1.5}Mn_{0.95}B_{0.05}O_4$, where $B$ = Ru, Co and Ni. It is shown that even a light doping is successful in suppressing the charge and orbital ordering found for the pure LSMO. In general, doping favours the carrier motion and, from a magnetic point of view, the set-up a spin-glass state. Moreover, structural parameters show an anisotropy in the lattice constant variation, with the tetragonal distortion increasing as the cell volume reduces, which may suggest a variation in the relative nature of the orbital character of the $e_g$ electrons along with the overall cation size.

*Corresponding Author: Dr. Lorenzo Malavasi, Dipartimento di Chimica Fisica "M. Rolla", INSTM, Università di Pavia, V.le Taramelli 16, I-27100, Pavia, Italy. Tel: +39-(0)382-987921 - Fax: +39-(0)382-987575 - E-mail: lorenzo.malavasi@unipv.it

# Introduction

In recent times there has been a revitalization of interest in the manganites as a result of the wide range of physical properties they display. In some of the doped compounds, there is an ordering of the charge carriers in defined orbitals, resulting in orbital ordering (OO) and charge ordering (CO). This is particularly true for the $n$ = 1 members of the Ruddlesden-Popper (R-P) series of manganites of general formula $A_{n+1}Mn_nO_{3n+1}$ in which $n$ 2D layers of $MnO_6$ corner-sharing octahedra are joined along the stacking direction and separated by rock-salt AO layers. Lowering the dimensionality of these systems makes compounds such as the $La_{1+x}Sr_{1-x}MnO_4$ (LSMO) solid solution an interesting model system for the study of the underlying doped $MnO_2$ planes and for a comparison with results for the double-layer ($n$ = 2) and perovskite ($n = \infty$) manganites.

In particular, in the $La_{0.5}Sr_{1.5}MnO_4$ compound, the Mn ions are nominally present in equal amounts in the +3 and +4 oxidation state and do order at around 240 K (CO), where it is believed that charge disproportionation of Mn ions occurs leading origin to two inequivalent site for $Mn^{3+}$ and $Mn^{4+}$ ions [1-3]. At temperatures lower than 110 K an anti-ferromagnetic (AF) order takes place where the spin of neighbouring Mn ions align antiferromagnetically within the $MnO_2$ planes. However, no trace of long-range magnetic order has been found in these compounds, where a spin-glass behaviour is instead observed, in accordance with the theoretical prediction of the absence of long-range magnetic order in one- or two-dimensional Heisenberg models [4].

It is interesting to note that the set-up of CO is accompanied by an OO of the $e_g$ orbitals. This OO seems to be the dominant "interaction" in $La_{0.5}Sr_{1.5}MnO_4$: recent resonant soft x-ray diffraction experiments at the Mn L edges have shown that the OO starts at the CO temperature and progressively increases by reducing the temperature and in addition it strongly increases at the Neél temperature [3]. Moreover, this study [3] revealed that the two major causes of the OO are the Jahn-Teller distortion and the short-range antiferromagnetic spin correlations. It is clear that the definition of the final ground state of $La_{0.5}Sr_{1.5}MnO_4$ is finely tuned by the several concurrent interactions (spin and orbital) that may occur in this manganite.

Foreign atoms doping may represent an interesting way to investigate the relative stability of these interactions. In this paper we are interested in looking at the role of low cation doping (5%) on the *B*-site of the structure. We then carried out the synthesis and structural, transport and magnetic characterization of $La_{0.5}Sr_{1.5}Mn_{0.95}B_{0.05}O_4$ samples, where *B* = Ru, Co and Ni. Among these ions, only Ru-doping was already studied on the $La_{0.5}Sr_{1.5}MnO_4$ compound, where for doping higher than 10% a ferromagnetic component coming from the Mn/Ru site has been observed [4].

All the data collected in this work have been compared to the properties of the pure $La_{0.5}Sr_{1.5}MnO_4$ manganite. Samples characterization was carried out by means of x-ray powder diffraction, DC transport measurement and magnetometry.

## Experimental Section

All the samples were synthesized through a wet-chemistry method defined as “propellant chemistry”. In this route, proper amounts of metal nitrates (all Aldrich ≥99.99+%) were used as oxidizers while a slight excess of urea (Carbonyldiamide, $CH_4N_2O$, Fluka >99.5%) was used as fuel. The oxidizers and fuel were dissolved in a small quantity of deionised water and heated on a hot-plate at 200°C until the whole solvent was evaporated. The final spongy powder was then calcined in a platinum crucible at 750°C for one hour. Finally, the powders were pressed and heated at 1300°C for 70 hours.

X-ray powder diffraction patterns were acquired on a “Bruker D8 Advance” diffractometer equipped with a Cu anode. Diffraction data were refined by means of the FULLPROFILE software [6]. Parameters refined were: zero shift, scale factor, lattice constants, atomic parameters, fractional occupancies and isotropic thermal factors.

Static magnetization was measured at 100 Oe from 350 K down to 2 K with a SQUID magnetometer (Quantum Design). *M vs. H* curves up to 7 tesla have been also collected (at 5K).

# Results

## *X-ray powder diffraction*

X-ray diffraction characterization of pure $La_{0.5}Sr_{1.5}MnO_4$ (LSMO) and $La_{0.5}Sr_{1.5}Mn_{0.95}Ru_{0.05}O_4$ (LSMRO) $La_{0.5}Sr_{1.5}Mn_{0.95}Co_{0.05}O_4$ (LSMCO), and $La_{0.5}Sr_{1.5}Mn_{0.95}Ni_{0.05}O_4$ (LSMNO), samples revealed the single-phase nature of the compounds which crystallize in the tetragonal *I*4/*mmm* space group.

Figure 1 reports, as a selected example, the Rietveld refined pattern of the pure LSMO compound. Structural parameters derived from the refinement of the patterns are listed in Table 1. Figure 2 reports the *a*, *c* lattice constants trend and cell volume and *c*/*a* parameter (inset) behaviour as a function of dopant ion. Here and in the following figures "Mn" represents the pure LSMO sample.

As can be appreciated from Figure 2 an anisotropy is present in the behaviour of the lattice parameters along the cation replacement: the *a* axis shrinks while the *c* axis expands going from Ru to Co dopant. Overall, the cell volume (see inset of Figure 2) reduces while the tetragonal distortion (*c*/*a*) increases passing from Ru to Co.

Figure 3 displays the trend of Mn-O bond lengths as a function of dopant ion. Two distinct Mn-O bonds are present in single-layered manganites: a longer out-of-plane (axial) bond and a shorter in-plane (equatorial) bond. The axial bond strongly reduces passing from LSMRO to LSMO and then remains practically constant. On the opposite, the in-plane bond progressively contracts passing from Ru-doped to Co-doped LSMO. Finally, an additional significant parameter, which gives an indication of the Mn-O octahedral distortion and in turn of the Jahn-Teller distortion, is the *D* parameter, defined as $D$ = Mn-$O_{apical}$/Mn-$O_{equatorial}$. Figure 4 shows the trend of *D* for the various dopants.

Let us discuss the variation of the structural parameters as a function of the dopant ions. The cell contraction passing from Ru-doped LSMO to the pure sample can rule out the presence of all Ru(V) since, for the same coordination, the ionic radii of this ion is 0.565 Å, while the "average" Mn size resulting from the +3.5 valence state is 0.587 Å. Since the ionic radius of Ru(IV) is 0.620 Å it is possible that Ru ions are present in a mixed $Ru^{4+}/Ru^{5+}$ valence state, as also suggested by previous Authors [5]. Both Ni- and Co-doped samples have smaller lattice volumes than pure LSMO. Based on the ionic radii for the various possible oxidation states and electronic configurations of Ni and Co, it can be concluded that nickel is present only as Ni(III) with a low-spin (LS) configuration ($t_{2g}^6\ e_g^1$) while we can not be conclusive on the Co state, since LS Co(III) and Co(IV) ions have very similar ionic radii (0.545 Å and 0.530 Å, respectively). Most probably, also the Co ions are present in a mixed valence state. However, divalent state for cobalt is hardly possible based on the lattice volume. This result is in contrast to what found in charge ordered perovskite manganites, *i.e.* the analogous $n$=∞ member of the R-P series of the LSMO single-layered manganite, where stable $Co^{2+}$ valence state was observed [7]. We also stress that the doping with aliovalent ions may induce a slight variation in the oxidation state of Mn ions as a consequence of charge compensation mechanism between dopant ions and manganese.

The trend shown in Figure 1 interestingly suggests that the unit cell does not expand in a isotropic way as the size of the ions increases. This in turn witnesses the presence of an electronic factor playing a role in the definition of the structural features of the samples.

From bond-length estimation it can be concluded that ruthenium is the only ions inducing an increase of the J-T distortion (see Figure 4) in the LSMO single-layered manganite. This is due to the stronger elongation of the out-of-plane Mn-O bond with respect to the in-plane Mn-O bond which indicates a strong stabilization of the $d_{3z^2-r^2}$ $e_g$ orbital. This may originate as a consequence of the reduction induced on the Mn-array by the Ru-doping and also from the nature of the magnetic interaction established by this ion which, opposite to Co and Ni, possess (?) more expanded 4*d* orbitals that may overlap more effectively with Mn-neighbour ions.

Finally, we note that for the Co and Ni-doped samples, the Mn-O array is not strongly affected by the substitution, as can be appreciated by looking at the variation of the Mn-O bond lengths. However, the lattice constants show a significant variation passing from LSMNO to LSMCO. In order to account for this also the La-O bond lengths have been calculated. For example, the expansion along the *c*-axis passing from Ni to Co is due to the concomitant elongation of the La-O(1) (*i.e.*, the one along the *c*-axis) and the La-Mn bonds. So, it looks that also the rock-salt layer is involved in the variation of the unit cell when a foreign atom is introduced in the lattice and part of the La-O bonds varies in an opposite way with respect to the Mn-O array in order to minimize the overall deformation of the structure.

## *Electrical transport properties*

On all the samples considered here, *i.e.* pure LSMO and the $La_{0.5}Sr_{1.5}Mn_{0.95}B_{0.05}O_4$ series ($B$ = Ni, Co and Ru) we performed low-temperature four-probe DC electrical conductivity measurements. Figure 5 reports the log $\sigma$ *vs.* $1/T$ plots obtained. Temperature ranges explored are in some cases reduced due to the very low $\sigma$-values of the samples, particularly at low-$T$.

Activation energies, calculated considering a purely activated transport according to $\rho = \rho_{\infty}\left( E_a \big/ KT \right)$, are reported in Table 2 together with the resistivity values at 260 K for all the samples.

Pure LSMO has an extremely high resistivity and presents a clear slope change around the CO/OO temperature (see inset of Figure 5). No clear sign of the AFM order is found in the $\sigma$ *vs.* $T$ curves. This is in agreement with the results found by other Authors [8, 9]. In all the cation doped samples there is a general increase of the electrical conductivity, particularly at low temperature, and the disappearance of slope change in the curves around the CO ordering. Moreover, cation doping is effective in promoting the charge carrier transport, as shown by the reduction of the activation energies. Note that for pure LSMO two different values of the activation energy are listed: the higher value refers to the data from RT to the slope change at around 240 K (CO/OO transition), while the lower value pertains to the data after this temperature.

Several effects must be considered in order to rationalise these data when the dopant ions are added in the pure LSMO: i) slight variation in the Mn valence state induced by charge compensation; ii) removal of charge and orbital ordering; iii) enhancement of FM interaction.

Ru-ions, as well as Ni and Co, give origin to an enhancement of the electron hopping by reducing the $E_a$. However, the resistivity at 260 K increases passing from LSMO to Ni-doped sample. This is probably connected to the pure paramagnetic behaviour found in the Ni-doped LSMO with respect to the other two samples where more extended short range magnetic interactions may constitute preferential conduction pathways. Most probably, the more expanded nature of the 4*d* orbitals of ruthenium is the origin of the "better" conductivity induced by this dopant with respect to Ni and Co.

## *Magnetic properties*

Figure 6 shows the molar susceptibility at 100 Oe for the samples studied in this work. In the inset is put in prominence the field-cooling (FC) and zero-field-cooling (ZFC) curves for the pure $La_{0.5}Sr_{1.5}MnO_4$.

Pure LSMO presents two clear transitions (marked with arrows in the inset of Figure 6): a first cusp at around 240 K (CO/OO transition) and a minimum around 110-120 K which marks the AFM transition. However, let us note that this AFM phase is not a long-range ordered phase but, overall, the sample behaves as a spin-glass. This is consistent with the theoretical prediction of the absence of long-range magnetic order in 2D Heisenberg systems. In particular the AFM phase found in this system extends as long-range only in the *a*-*b* plane, while it has a finite correlation length perpendicular to the $MnO_2$ planes [9].

Doped samples do not present any sign of the magnetic transitions found in the pure sample. In detail, Ni-doped LSMO behaves as a pure paramagnet for temperature up to ~25 K, while for lower temperatures the FC and ZFC curves deviate one from each other. This implies the evolution of a magnetic order. An analogous trend is found for both the Ru- and Co-doped LSMO. However, for these two samples the set-up of magnetic interactions start at higher temperatures with respect to the Ni-doped LSMO and in particular a separation between the FC and ZFC curves for the $La_{0.5}Sr_{1.5}Mn_{0.95}Co_{0.05}O_4$ is already present at room temperature. Let us note that the shape of the susceptibility curves for all the doped samples resembles that of a spin-glass, particularly for the presence of a distinct maximum in the ZFC. Interestingly, the peak in the ZFC curves, which can be thought as the freezing transition to the low temperature phase with random alignment of the spins, falls at nearly the same temperature, ~22 K, for all the samples, thus suggesting the presence of the same type of magnetic interaction for all of them even though with different strength from sample to sample. The origin of the magnetic frustration in these samples is probably due to the presence of competing FM and AFM interactions and also from the disorder induced by the cation doping.

Finally, Figure 7 presents the field dependence of the magnetization at 5 K. Only for pure LSMO there is an almost linear dependence of *M* with the field. For all the other samples a marked curvature occurs. However, according to the short-range nature of the magnetic interaction, the magnetization never saturates for all the samples. Qualitatively we may state that in the LSMRO sample the strength of the magnetic interaction seems to be higher with respect to the other two doped samples.

# Conclusion

In this paper we reported the results of an investigation which aimed to study the role of cation doping on the *B*-site (Mn) of the single layered $La_{0.5}Sr_{1.5}MnO_4$ manganite. The main results we obtained can be summarized in the following:

1. We successfully synthesised single-phase tetragonal $La_{0.5}Sr_{1.5}Mn_{0.95}Ru_{0.05}O_4$, $La_{0.5}Sr_{1.5}Mn_{0.95}Co_{0.05}O_4$, and $La_{0.5}Sr_{1.5}Mn_{0.95}Ni_{0.05}O_4$ samples. The last two have never been prepared and characterized before. In addition, we employed for the preparation of these compounds, for the first time, a wet-chemistry method that was previously employed for perovskite manganites [10];
2. Structural parameters show an anisotropy in the lattice constant variation, with the tetragonal distortion increasing as the cell volume reduces; this may suggest a variation in the relative nature of the orbital character of the $e_g$ electrons along with the overall cation size;
3. CO/OO and AFM transitions characterizing the pure LSMO disappear in all the doped samples, thus suggesting that a light doping, as the one realized here, is already able to destroy the orbital ordered ground state of the $La_{0.5}Sr_{1.5}MnO_4$ manganite;
4. Cation doping gives origin to an easier carrier motion in the samples as shown by the reduction of the activation energies;
5. Finally, from a magnetic point of view the doped samples behave as spin-glass with a common freezing temperature for all of them.

# Acknowledgement

Financial support from the Italian Ministry of Scientific Research (MIUR) by PRIN Projects (2004) is gratefully acknowledged. One of us (L.M.) gratefully acknowledges the financial support of the "Accademia Nazionale dei Lincei".

## Figures Captions

**Figure 1** –Refined x-ray diffraction pattern of LSMO. Red crosses represent the experimental pattern, the black line is the calculated one, while the blue line is the difference between them. Bragg peaks appear as vertical green lines.

**Figure 2** – *a*, *c* lattice constants, cell volume and *c*/*a* parameter (inset) for the different cation doped LSMO samples.

**Figure 3** – Mn-O axial and equatorial bonds for the different cation doped LSMO samples.

**Figure 4** – *D* parameter for the different cation doped LSMO samples.

**Figure 5** – Logarithm of conductivity ($\sigma$) *vs.* $1/T$ for the $La_{0.5}Sr_{1.5}Mn_{0.95}B_{0.05}O_4$ samples where $B$ = Ru, Mn (pure sample), Ni and Co. In the inset it is reported the Logarithm of $\sigma$ *vs.* $1/T$ for $La_{0.5}Sr_{1.5}MnO_4$ with the arrow highlighting the slope change around the CO temperature.

**Figure 6** – Molar susceptibility *vs.* $T$ for the $La_{0.5}Sr_{1.5}Mn_{0.95}B_{0.05}O_4$ samples where $B$ = Ru, Mn (pure sample, in the inset), Ni and Co.

**Figure 7** – Magnetization *vs.* $H$ for the $La_{0.5}Sr_{1.5}Mn_{0.95}B_{0.05}O_4$ samples, $B$ = Ru, Mn, Ni and Co.

## Table Caption

**Table 1** Structural parameters derived from the x-ray diffraction for the $La_{0.5}Sr_{1.5}Mn_{0.95}B_{0.05}O_4$ where $B$ = Ru, Mn (pure sample), Ni and Co.

**Table 2**. Activation energies and $\rho$-values at 260 K for the samples considered.

# Table 1

| | ***Ru*** | ***Mn*** | ***Ni*** | ***Co*** |
|---|---|---|---|---|
| $a$ (Å) | 3.86478(4) | 3.86231(9) | 3.85883(5) | 3.85810(4) |
| $c$ (Å) | 12.4241(1) | 12.4292(2) | 12.4292(2) | 12.4311(1) |
| $V$ (Å$^3$) | 185.574(3) | 185.077(4) | 185.079(4) | 185.692(3) |
| $c/a$ | 3.21469(1) | 3.2209(2) | 3.2210(2) | 3.2221(1) |

# Table 2

| ***Sample*** | $E_{att}$ (meV) | $\rho_{260\,K}$ (Ωcm) |
|---|---|---|
| $La_{0.5}Sr_{1.5}MnO_4$ | 357-443 | 11.7 |
| $La_{0.5}Sr_{1.5}Mn_{0.95}Ru_{0.05}O_4$ | 135 | 27.3 |
| $La_{0.5}Sr_{1.5}Mn_{0.95}Co_{0.05}O_4$ | 200 | 92.5 |
| $La_{0.5}Sr_{1.5}Mn_{0.95}Ni_{0.05}O_4$ | 154 | 265.1 |

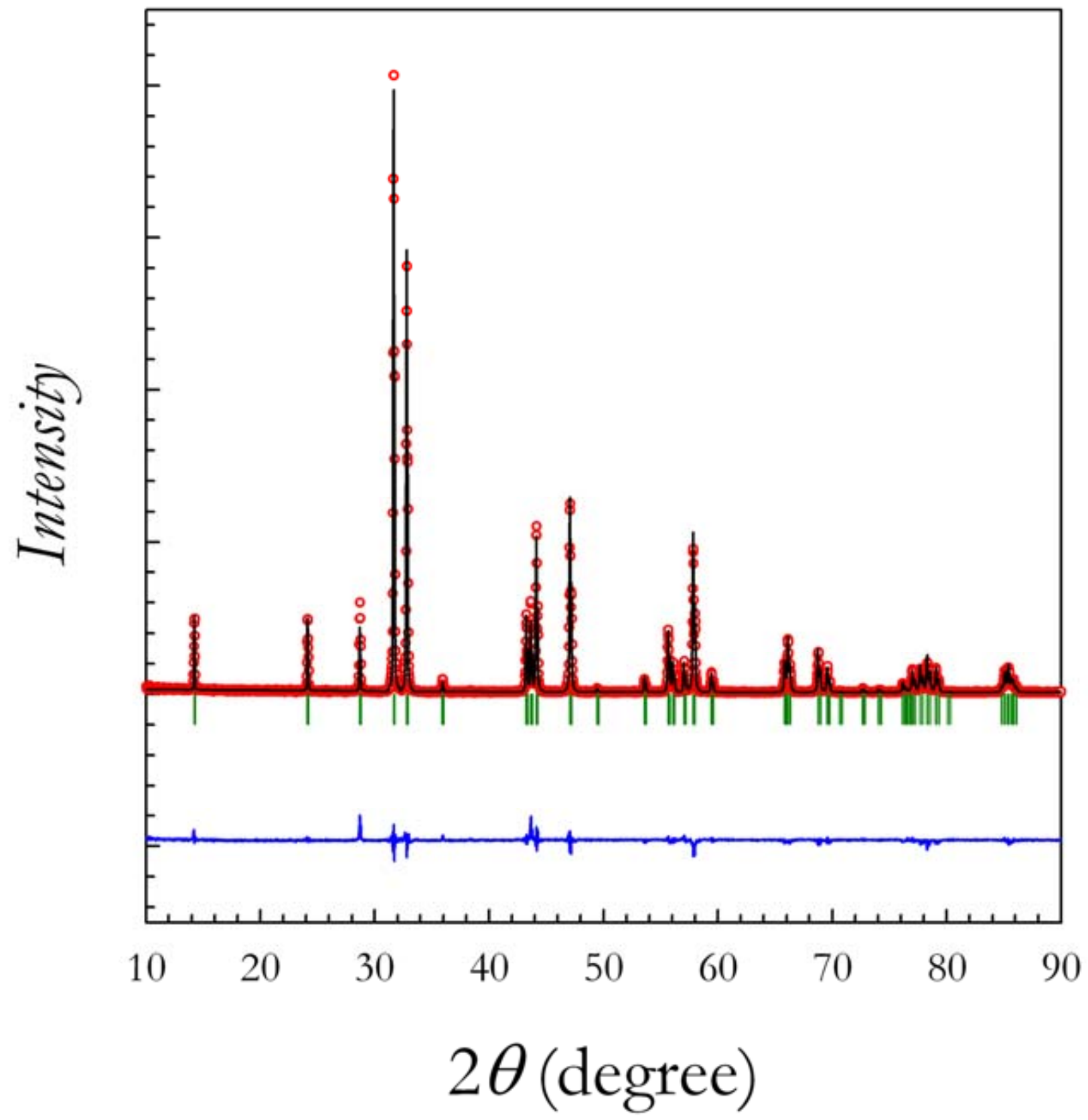

**Figure 1**

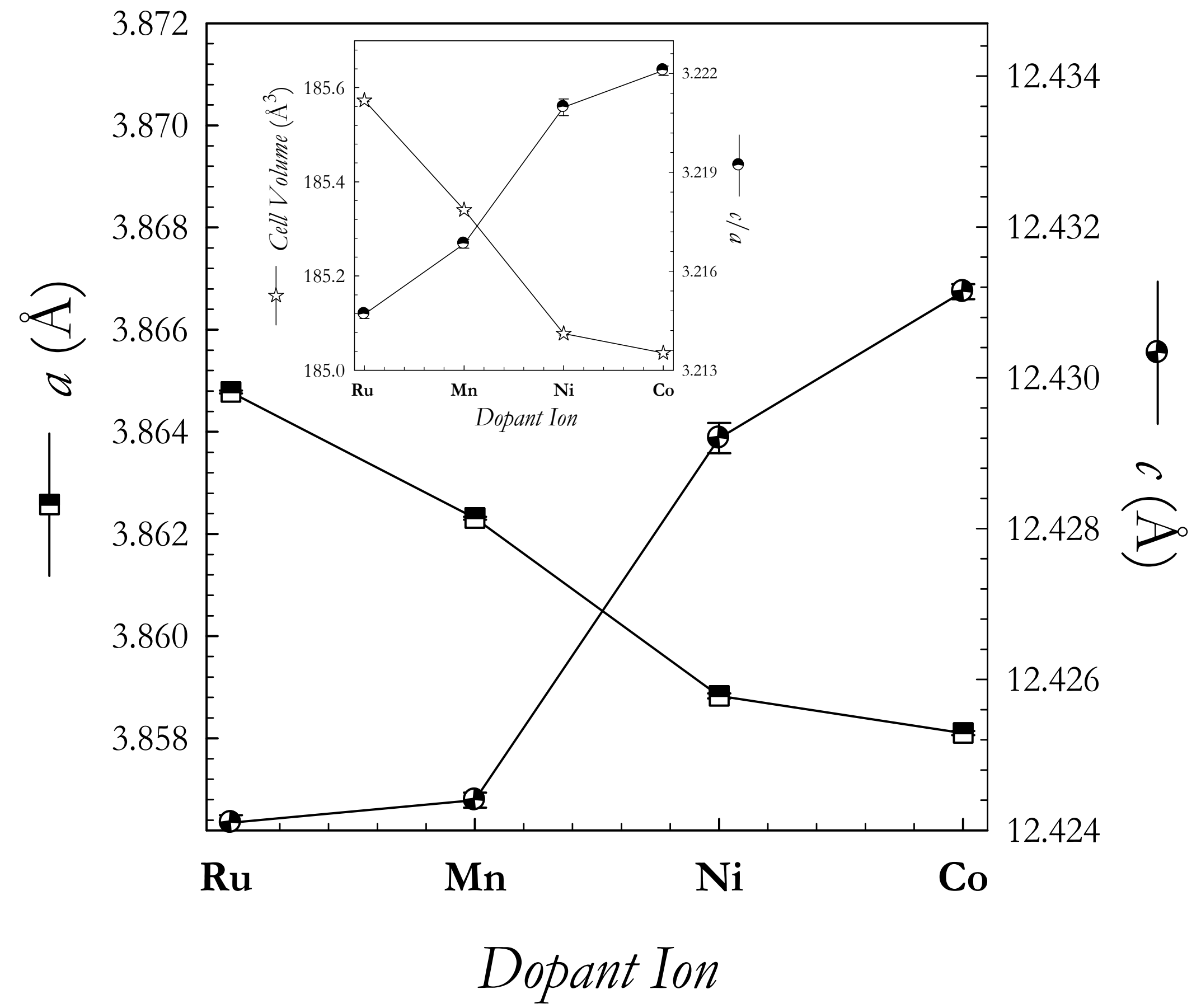

**Figure 2**

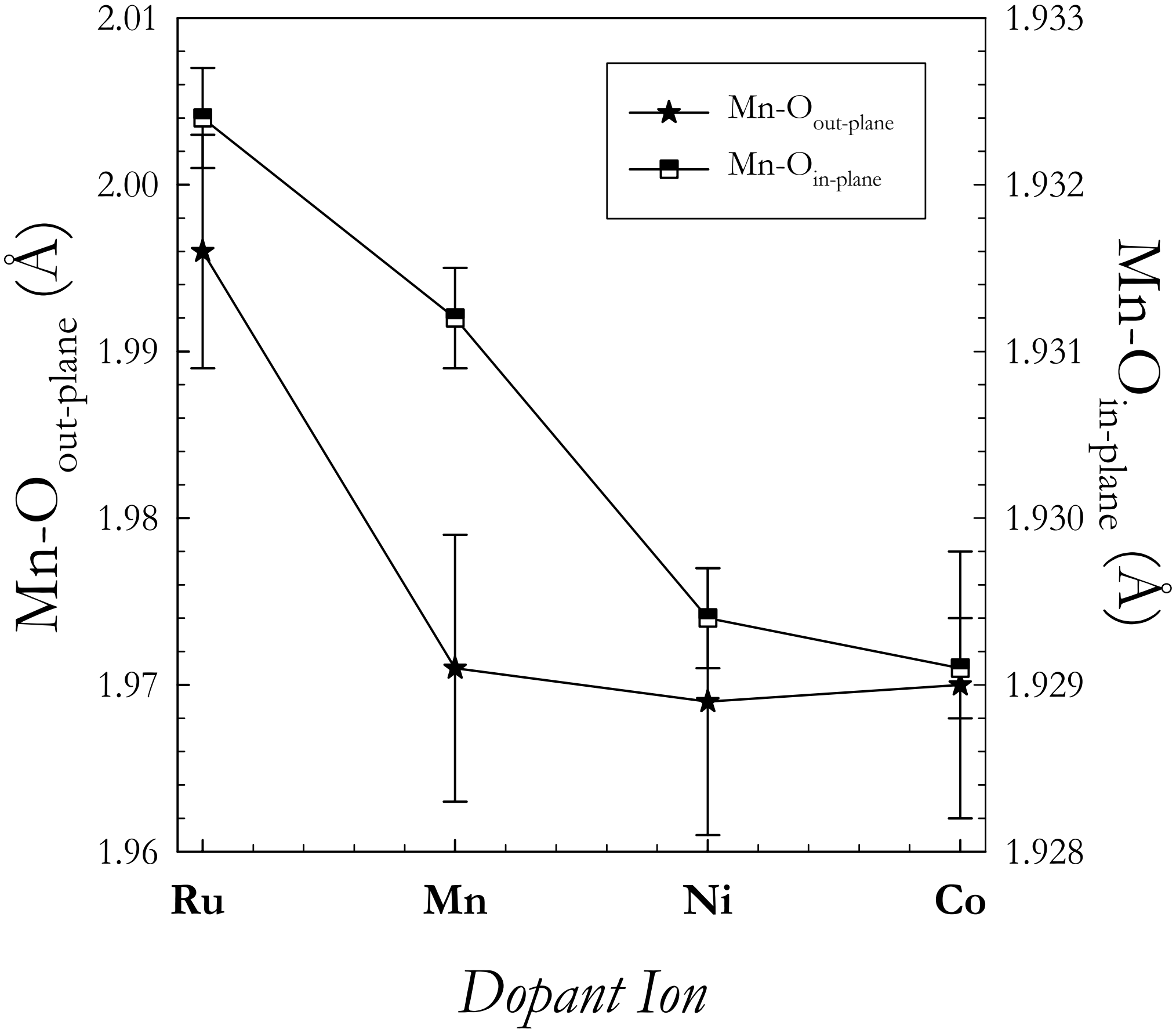

**Figure 3**

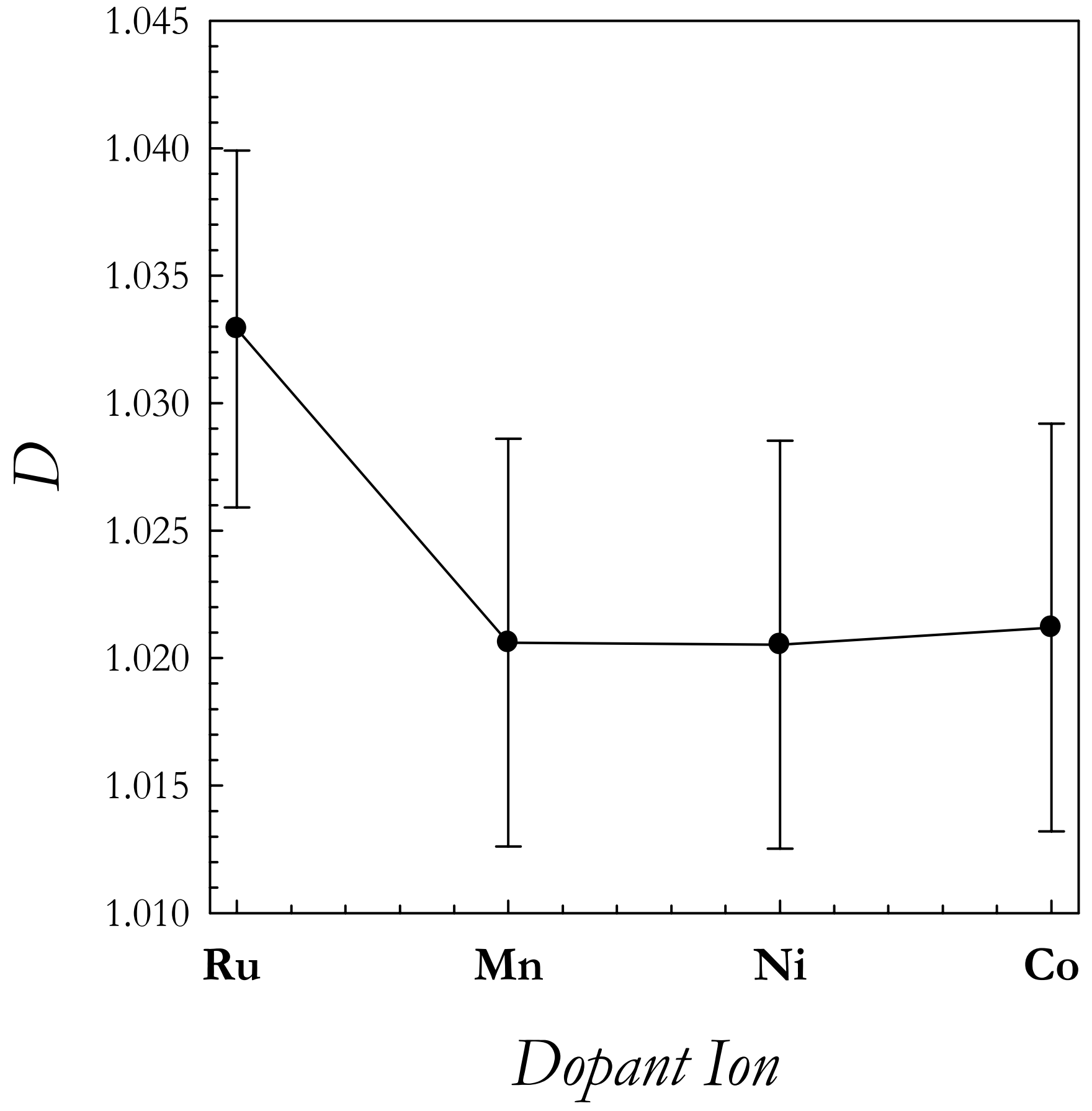

**Figure 4**

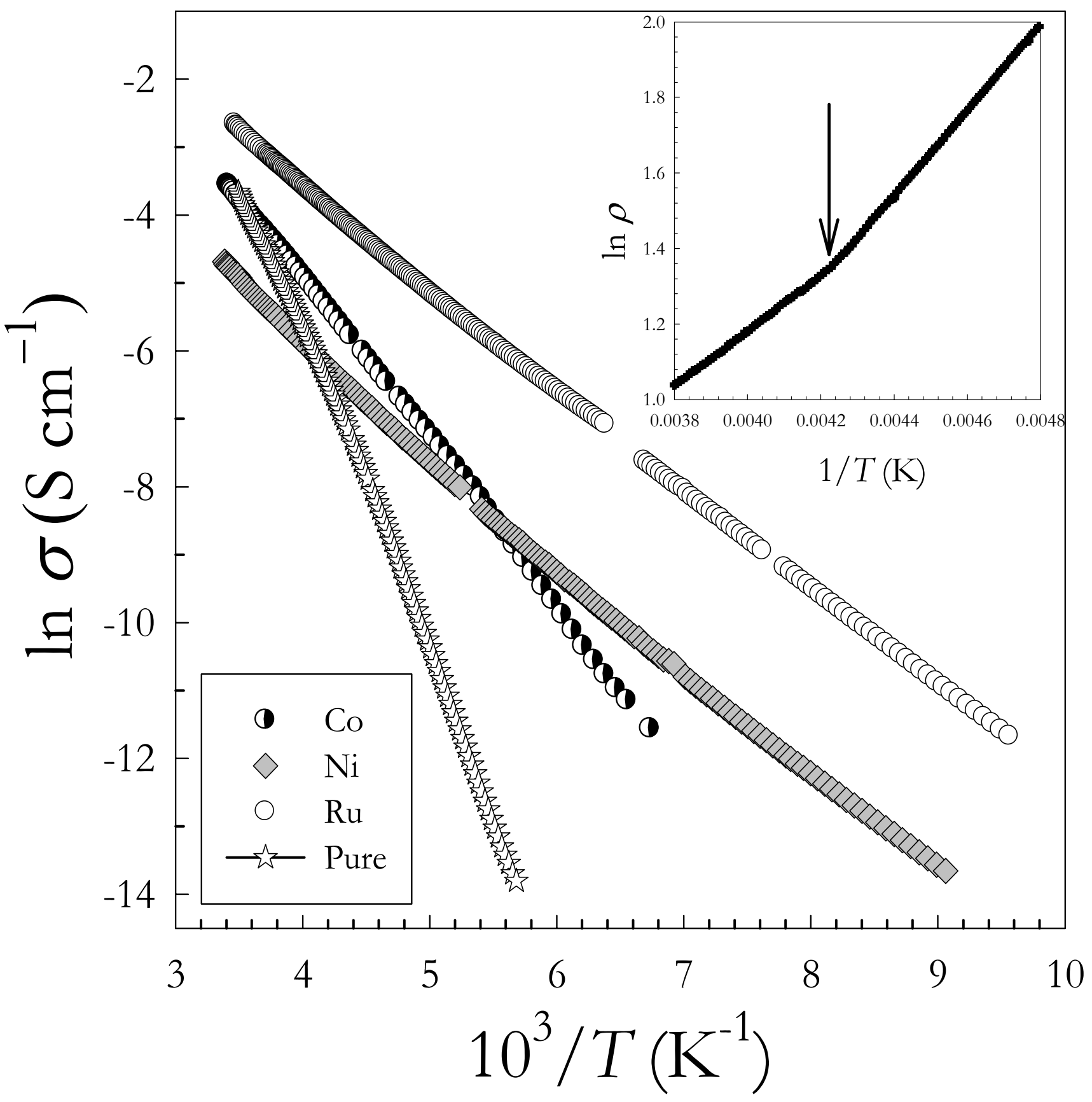

**Figure 5**

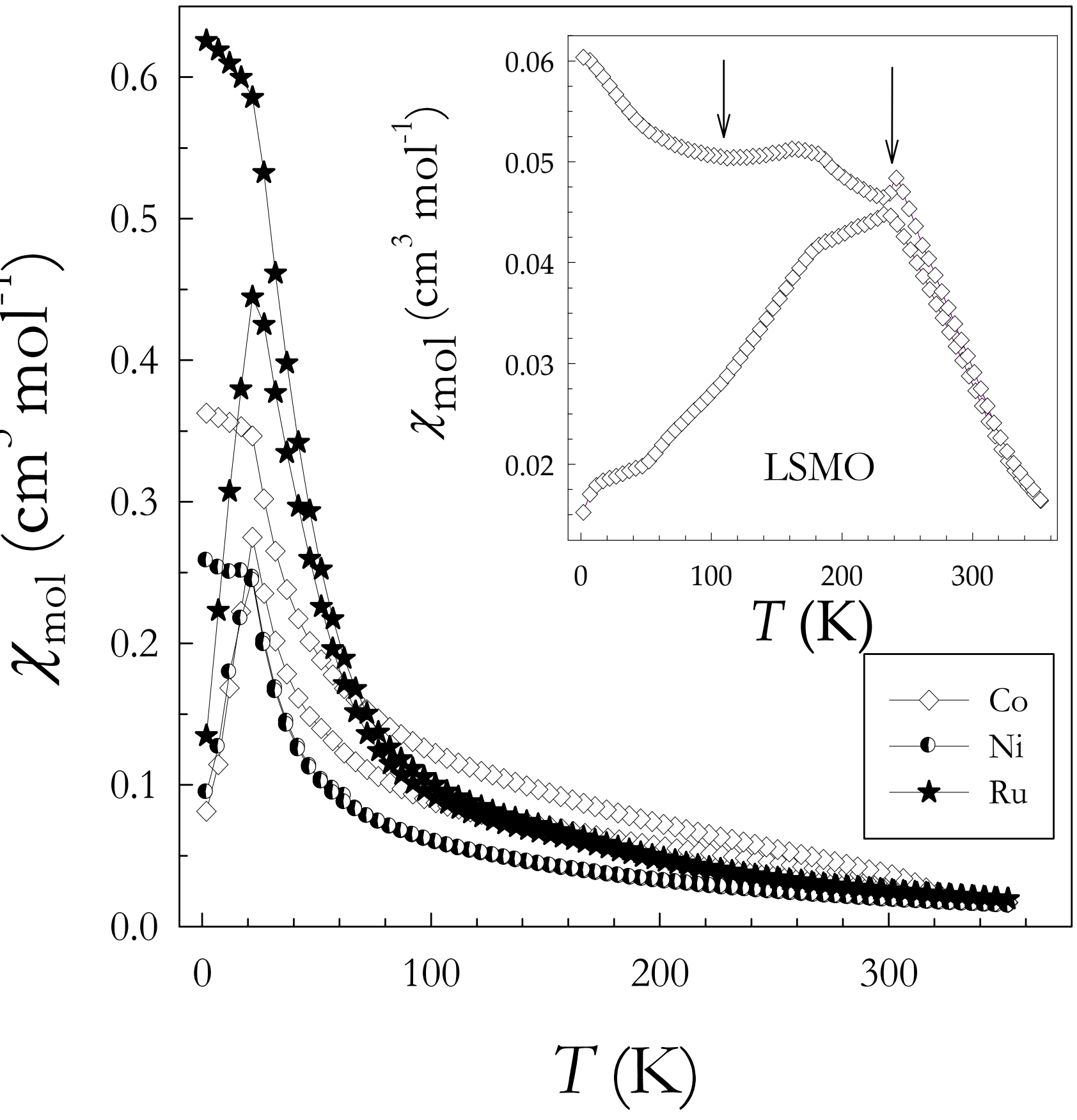

**Figure 6**

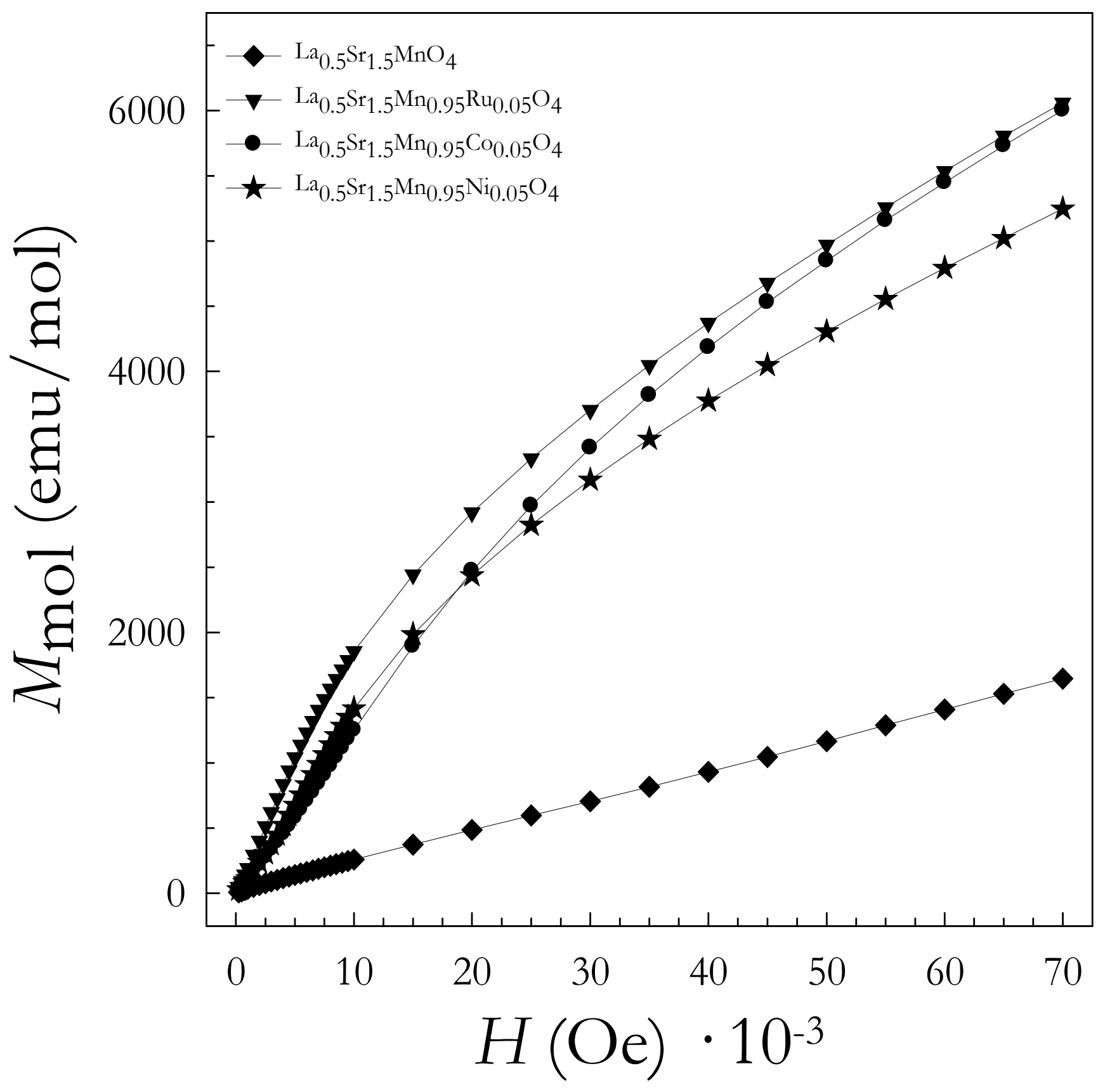

**Figure 7**